\documentclass[a4paper]{aa} 
\usepackage{graphicx,float}  
  
\def \xmm {XMM-Newton}  
\def \src {GX\thinspace13$+$1}  
\def \degmark{^\circ}  
\def \nh {N${\rm _H}$}  
  
\def \hcm {\hbox {\ifmmode $ atom cm$^{-2}\else atom cm$^{-2}$\fi}} 
  
\def \arcsec {\hbox{$^{\prime\prime}$}}  
\def \chisq {$\chi ^{2}$}  
\def \rchisq {$\chi_{\nu} ^{2}$}  
\def\approxgt{\mathrel{\hbox{\rlap{\lower.55ex \hbox {$\sim$}}  
        \kern-.3em \raise.4ex \hbox{$>$}}}}  
\def\approxlt{\mathrel{\hbox{\rlap{\lower.55ex \hbox {$\sim$}}  
        \kern-.3em \raise.4ex \hbox{$<$}}}}  
\newcommand{\mc}{\multicolumn}

\begin{document}

\title{Discovery of complex narrow X--ray absorption features   
from the low-mass X--ray binary \src\ with XMM-Newton}  
  
\author{L. Sidoli\inst{1}  
	\and A. N. Parmar\inst{2}  
        \and T. Oosterbroek\inst{2}  
         \and D. Lumb\inst{2}  
}  
\offprints{L. Sidoli, \email{sidoli@ifctr.mi.cnr.it}}  
  
\institute{Istituto di Fisica Cosmica ``G. Occhialini", CNR,  
	via Bassini 15, I-20133 Milano, Italy  
	\and  
       Astrophysics Division, Research and Scientific Support   
       Department of ESA, ESTEC,  
       Postbus 299, NL-2200 AG Noordwijk, The Netherlands  
}  
\date{Received 19 December 2001  / Accepted: 31 January 2002}  
  
\authorrunning{L. Sidoli et al.}  
  
\titlerunning{XMM-Newton observations of \src}  
  
\abstract{We report the detection of a complex of 
narrow X--ray absorption  
features from the low-mass X--ray binary \src\ during  
3 XMM-Newton observations in 2000 March and April.   
The features are consistent with being due to resonant scattering 
of the K$\alpha$ and K$\beta$ lines of He- and H-like iron  
(Fe\,{\sc xxv} and Fe\,{\sc xxvi}) and H-like calcium 
(Ca\,{\sc xx}) K$\alpha$.  Only the Fe\,{\sc xxvi} K$\alpha$   
line has been previously observed from \src.  
Due to the closeness in energy the Fe\,{\sc xxv} and   
Fe\,{\sc xxvi} K$\beta$ features may also be ascribed to   
Ni\,{\sc xxvii} and Ni\,{\sc xxviii} K$\alpha$, respectively.  
We also find evidence for the presence of a   
deep ($\tau \sim 0.2$) Fe~{\sc xxv} absorption edge 
at 8.83~keV. The fits also require the presence of a broad 
emission feature whose energy and width are poorly determined, 
partly due to the presence of the deep Fe K$\alpha$ 
features which severely cut into the feature and partly due 
to fit differences when using different XMM-Newton instruments. 
The equivalent widths of the lines do not show  
any obvious variation on a timescale of a few days suggesting  
that the absorbing material is a stable feature of the 
system and present during a range of orbital phases.   
  } \maketitle  
  
\section{Introduction}  
\label{sect:intro}

The bright and persistent low-mass X--ray binary (LMXRB) \src\   
(4U\thinspace1811--17)  
is an X--ray burst source (Fleischman \cite{f:85}; Matsuba \cite{m:95})  
with radio (Grindlay \& Seaquist \cite{gs:86})   
and infrared counterparts  
(Naylor et al. \cite{ncl:91}; Charles \& Naylor \cite{cn:92}).   
It displays a 15\% modulation in X--ray intensity with a period   
of 24.7$\pm$1~days. The X--ray intensity is anti-correlated   
with the hardness ratio (Corbet \cite{c:96}).  
Recent estimates of the spectral type of the companion (an evolved   
late-type K5~{\sc iii} star; Bandyopadhyay et al. \cite{b:99}),   
together with the binary system parameters are consistent with   
this modulation being the orbital period, although further confirmation  
is required   (Bandyopadhyay  et al. \cite{b:02}).
\src\ has been classified as an atoll source, but in some respects its 
properties are more similar to those of the Z-sources   
(Homan et al. \cite{h:98}; Schulz et al. \cite{s:89};   
Bandyopadhyay et al. \cite{b:99}),   
having persistent radio emission    
at a similar level to the Z-sources (Fender \& Hendry \cite{fh:00}),   
and a high accretion rate of $\sim$0.1 Eddington.   
A quasi-periodic oscillation (QPO) was discovered   
at 57--69~Hz by Homan et al. (\cite{h:98}).  
Distance estimates for \src\ range from 4.7 to 7.0~kpc   
(Christian \& Swank \cite{cs:97}).  
Infrared spectroscopy favors a distance of 7$\pm$1~kpc   
(Bandyopadhyay et al. \cite{b:99}).

\begin{figure*}
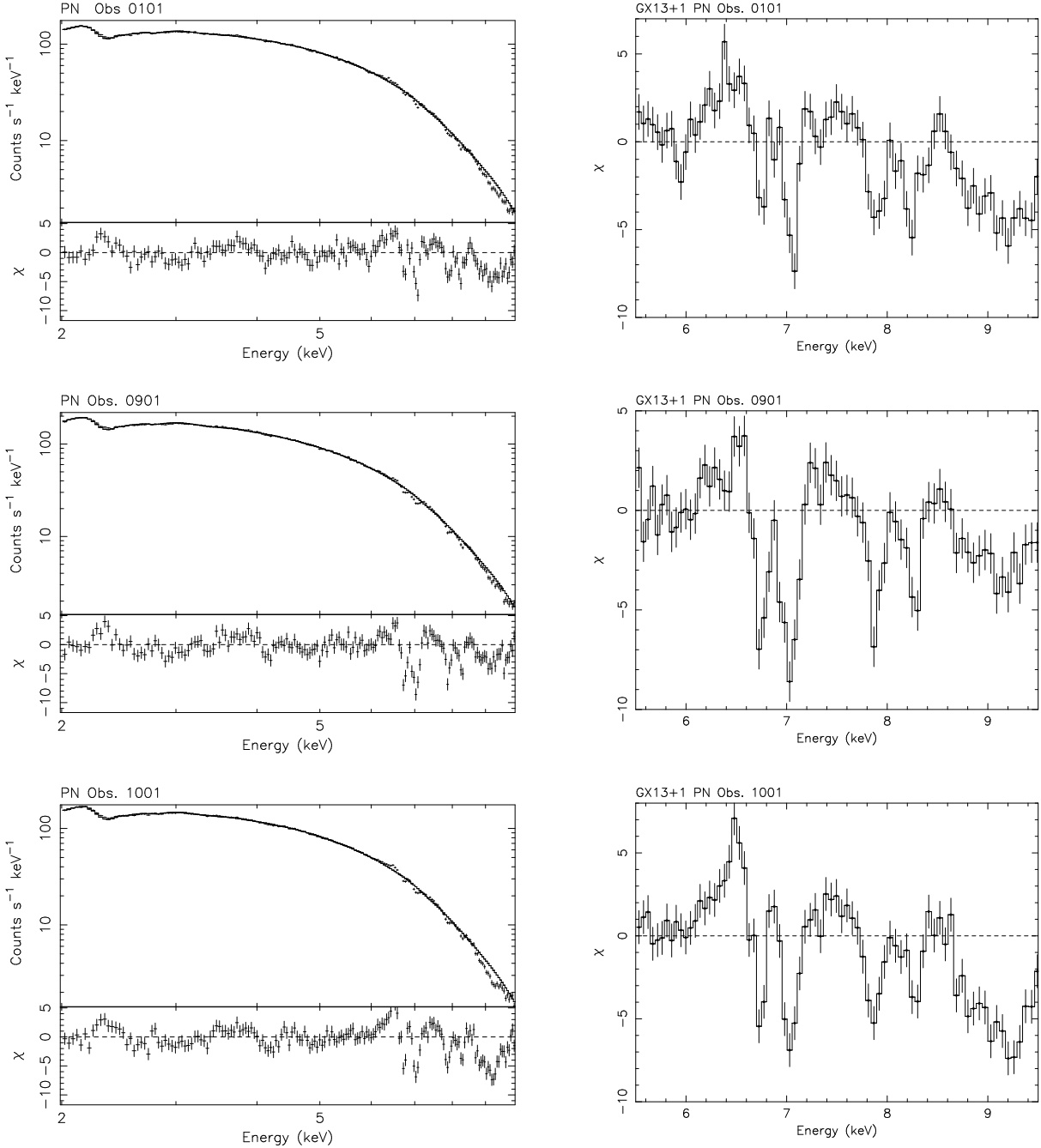
  
\hbox{\hspace{0.5cm}  
\includegraphics[height=7.9cm,angle=-90]{h3372f1a.ps}  
\hspace{1.0cm}  
\includegraphics[height=6.9cm,angle=-90]{h3372f1b.ps}}  
\vbox{\vspace{0.1cm}}  
  
\hbox{\hspace{0.5cm}  
\includegraphics[height=7.9cm,angle=-90]{h3372f1c.ps}  
\hspace{1.0cm}  
\includegraphics[height=6.9cm,angle=-90]{h3372f1d.ps}}  
\vbox{\vspace{0.1cm}}  
   
\hbox{\hspace{0.5cm}  
\includegraphics[height=7.9cm,angle=-90]{h3372f1e.ps}  
\hspace{1.0cm}  
\includegraphics[height=6.9cm,angle=-90]{h3372f1f.ps}}

\caption[]{The EPIC PN spectra for the 3 on-axis observations of \src. 
In the left panels, the best-fit continuum model consisting of 
absorbed multi-color blackbody and blackbody components, together with 
a broad emission feature at 6.8~keV is shown with the normalizations 
of the narrow discrete absorption features, and the optical depth of 
the absorption edge, set to zero. On the right, expanded views of the 
residuals (in units of standard deviation) in the 5.5--9.5~keV energy 
range. The four absorption lines, a broad emission line at 6.8~keV 
which is strongly affected by the two strongest absorption features 
and the absorption edge at 8.83~keV are clearly visible} 
\label{fig:spec}  
\end{figure*}

\begin{table*}  
\caption{\xmm\ on-axis observation log of \src.  
All observations used the medium filter}  
\label{tab:log}  
\begin{tabular}[c]{lrrccll}  
\hline\noalign{\smallskip}  
Rev. & Obs. ID&Inst. &  Start time (2000)    & \mc{1}{c}{Net Exp.}    
& Mode  \\  
      &  &  EPIC      & (dy~mon~hr:mn)   & \mc{1}{c}{(ks)}  &    \\  
\noalign{\smallskip\hrule\smallskip}  
0056 & 0101 & MOS1 &  30 Mar 15:13 &  5.9      &  Full Frame  \\  
     &      & PN   &  30 Mar 14:47 &  2.5      &  Timing & \\  
0057 & 0901 & MOS1 &  01 Apr 06:28 &  4.9      &  Full Frame  \\  
     &      & PN   &  01 Apr 05:52 &  2.1      &  Timing \\  
0057 & 1001 & MOS1 &  01 Apr 09:54 &  4.6      &  Full Frame \\  
     &      & PN   &  01 Apr 09:28 &  2.3      &  Timing \\  
\noalign{\smallskip\hrule\smallskip}  
\end{tabular}  
\end{table*}

A spectral survey of Fe-K emission  
from 20 LMXRB with ASCA is reported in Asai et al.~(\cite{a:00}).   
Significant iron emission was detected from \src\ 
with a line centroid of $6.44\pm 0.05$~keV, 
an equivalent  
width (EW) of $37 \, ^{+10} _{-8}$~eV and a line width $\sigma$$\sim$130$\pm$40~eV.  
The X--ray continuum could be modeled with a two-component model,   
consisting  
of a multicolor disk-blackbody with a temperature, $kT$,   at the
inner disk radius of 0.76~keV  
and a blackbody with a $kT$ of 1.34~keV, absorbed by  
a column density, $N_{\rm H}$, of $2.9 \, 10^{22}$~atom~cm$^{-2}$.  
The source flux (1--10~keV, corrected for absorption)   
was 1.2$\,$10$^{-8}$~erg~cm$^{-2}$~s$^{-1}$, which translates into  
a luminosity of $7\,$10$^{37}$~erg~s$^{-1}$ (at 7~kpc).  
Recently, Ueda et al. (\cite{u:01}) re-analyzed the ASCA  
Solid-State Imaging Spectrometer (SIS) observation of \src\ 
and discovered that the spectral region around Fe is more complex 
than had been realized. Their fits required the presence of a  
narrow ($\sigma < 70$~eV) absorption line   
at $7.01 \pm 0.03$~keV with an EW of 35$\pm$8~eV, which they interpreted as   
resonant scattering of the K$\alpha$ line from   
Fe\,{\sc xxvi} ions with a column density $\approxgt$$10^{18}$~cm$^{-2}$, 
a narrow ($\sigma < 220$~eV) emission line at $6.42 \pm 0.08$~keV with 
an EW of $19 \pm 8$~eV and an absorption edge at $7.61 \pm 0.13$~keV 
with an optical depth of $0.13 \pm 0.05$.   
This was the first detection of an Fe absorption line in an X--ray binary   
known to contain a neutron star.   
Previously, the only X--ray binaries known to exhibit such Fe absorption   
features were two superluminal jet sources   
GRO\,J1655-40 (Ueda et al.~\cite{u:98};   
Yamaoka et al.~\cite{y:01}) and GRS\,1915+105 (Kotani et al.~\cite{k:00};  
Lee et al.~\cite{l:01}). As Ueda et al.~(\cite{u:01}) comment, the 
interpretation of the edge is not simple and it could be a combination 
of an absorption edge from moderately ionized Fe together with line 
absorption from a number of prominent K$\alpha$ or K$\beta$  
lines of Fe and Ni.  
  
XMM-Newton is significantly improving our understanding   
of the LMXRB emission processes, accretion geometries and the physical   
conditions of the material through which the   
X--ray emission propagates.  
Reflection Grating Spectrometer    
(RGS) observations of the eclipsing and dipping LMXRB EXO\thinspace0748-676   
(Cottam et al.~\cite{c:01})   
revealed  several discrete absorption and emission   
features from ionized neon  
(Ne\,{\sc x}~Ly$\alpha$, Ne\,{\sc ix}~Ly$\alpha$),   
oxygen (O\,{\sc viii} Ly$\alpha$, O\,{\sc vii} He-like complex)   
and nitrogen (N\,{\sc vii}~Ly$\alpha$).  
These line features are broadened  
(velocity widths of 1000--3000~km~s$^{-1}$),   
but with no evidence for a velocity shift ($<$300~km~s$^{-1}$).  
No obvious dependence on orbital phase was found.  
  
During an XMM-Newton observation of the eclipsing and   
dipping LMXRB MXB\thinspace1659--298  
resonant absorption features identified   
with O~{\sc viii}~1s-2p, 1s-3p and   
1s-4p, Ne~{\sc x} 1s-2p, Fe~{\sc xxv}~1s-2p,   
and Fe~{\sc xxvi}~1s-2p transitions,  
together with a broad Fe emission feature at $6.47 ^{+0.18} _{-0.14}$~keV   
were discovered (Sidoli et al.~\cite{si:01}).   
The line widths are unresolved in the RGS and correspond to  
velocities of $<$600~km~s$^{-1}$ for O~{\sc viii}~1s-2p and 
$<$2300~km~s$^{-1}$ for the Fe~{\sc xxvi} K$\alpha$ line 
observed using the European Photon Imaging Camera (EPIC). 
The absorption  
features occur through a wide range of orbital phases,   
so the absorbing material  
is most likely located above, or below, the accretion disk,   
in a cylindrical geometry with an axis   
perpendicular to that of the disk.  
  
Similar absorption features have been reported by  
Parmar et al. (\cite{p:01})   
during an XMM-Newton observation of the LMXRB X\thinspace1624$-$490,  
sometimes called the ``Big Dipper''.   
Here the lines   
are identified with the resonant K$\alpha$ absorption of Fe\,{\sc xxv}   
and Fe\,{\sc xxvi}, with   
Ni\,{\sc xxvii} K$\alpha$ and Fe\,{\sc xxvi} K$\beta$ possibly detected.   
Again, no obvious dependence on orbital phase was observed, except  
during an absorption dip.  
Here, we report the discovery  
of discrete X--ray absorption features from highly  
ionized Fe and Ca (and possibly Ni) in XMM-Newton EPIC spectra of \src.    
 
\section{Observations}  
\label{sect:obs}  
   
The XMM-Newton Observatory (Jansen et al. \cite{j:01}) includes three  
1500~cm$^2$ X--ray telescopes each with an EPIC at the focus.   
Two of the EPIC imaging spectrometers use MOS CCDs (Turner et al.  
\cite{t:01}) and one uses a PN CCD (Str\"uder et al. \cite{st:01}).  
The region of sky containing \src\ was observed by XMM-Newton   
a number of times as part of the calibration programme.  
Three on-axis observations were performed (see Table~1).  
Since an analysis of two off-axis observations 
(2000 April 1 13:34 and April 3 13:34 UTC)
give similar spectral results,   
we discuss only the better calibrated on-axis observations.  
In order to minimize the effects of pile-up the PN and MOS2  
were operated in their Timing Modes.   
The MOS1 CCD was operated in its normal  
Full Frame Mode and the data are strongly affected by pile-up 
  (when two photons are detected closely in time and the 
energy is incorrectly assigned to be the sum of the
individual energies; see Str\"uder et al.~\cite{st:01}).   
The effects of pile-up in MOS1 were minimized by extracting events in an  
annulus outside the 30\arcsec\ radius  
core of the \src\ point spread function (PSF). Since there are 
far fewer MOS events than PN ones, the MOS1 data are only used to 
verify the overall shape and energy calibration of the PN data. 
  
Raw data products were extracted from  
the public XMM-Newton archive and then reprocessed using  
version 5.2 of the Science Analysis Software (SAS),  
before being further filtered using {\sc xmmselect}.   
Only X--ray events corresponding to patterns 0 (single pixel  
events) were selected.  
Background spectra were obtained from source free regions taken from  
the same observations.  
For the Timing Mode observations, appropriate public response matrices   
are not yet available.  
Thus, the standard PN matrix for Full Frame Mode has been adopted.  
This slightly affects the modeling of the continuum, and possibly the 
energies of the lines (F.~Haberl, priv. communication),  
but is unlikely to affect estimates of their EWs.  
We concentrate here on the analysis of the narrow absorption 
features evident in the spectra and defer a detailed discussion 
of the continuum properties to a later paper.

\begin{table*}  
\begin{center}  
\caption[]{Parameters of the absorption lines detected in the XMM-Newton  
spectra of \src. A continuum model consisting of absorbed blackbody and disk-blackbody   
components was used}  
\begin{tabular}{ll|ll|ll|ll}  
\hline  
\noalign {\smallskip}  
Component & Parameter              &  \mc{2}{c}{Obs.~0101}  &   	  
  \mc{2}{c}{Obs.~0901}     	&  \mc{2}{c}{Obs.~1001}        \\  
	  & 	                   &  MOS1       &  PN 	&  MOS1       &  PN  	&  MOS1       &  PN  \\  
\noalign {\smallskip}  
\hline  
\noalign {\smallskip}                          			 		  
Ca\,{\sc xx} abs	& $E_{{\rm line}}$ (keV) 		& $4.15 \, ^{+0.05}_{-0.05}$	& $4.15 \, ^{+0.3}_{-0.05}$	& $4.16 \, ^{+0.03}_{-0.03}$	&  $4.18 \, ^{+0.12}_{-0.35}$ 	& $4.13 \, ^{+0.04}_{-0.03}$		&   $4.19 \, ^{+0.04}_{-0.05}$	 \\  
line			&$\sigma$ (keV)           		& $<$0.05 			&   $<$0.1 			&   $<$0.1 			&  $<$0.2 			&   $<$0.05 				&   $<$0.1  \\  
			&EW         (eV)    	 		& $-7.6 \, ^{+7.1} _{-7.6}$	& $-3.3 \, ^{+1.9} _{-2.1}$	& $-11.3 \, ^{+4.3} _{-7.0}$	&  $-4.3 \, ^{+2.1} _{-2.3}$ 	& $-7.7 \, ^{+3.8} _{-5.8}$		&   $-6.0 \, ^{+2.6} _{-2.9}$	 \\  
Fe\,{\sc xxv} K$\alpha$ abs	& $E_{{\rm line}}$ (keV) 	& $6.73 \, ^{+0.04}_{-0.04}$ 	& $6.75 \, ^{+0.10}_{-0.02}$ 	& $6.76 \, ^{+0.05 }_{-0.05}$	&  $6.75 \, ^{+0.01}_{-0.02}$ 	& $6.72 \, ^{+0.08}_{-0.06}$		&   $6.72  \, ^{+0.02} _{-0.01}$ \\  
line			&$\sigma$ (keV)           		&   $<$0.1 			&   $<$0.1 			&   $<$0.1 			&  $<$0.1 			&   $<$0.05 				&   $<$0.04  \\  
			&EW         (eV)    	 		& $-23.2 \, ^{+13.2} _{-19.2}$	&   $-23.2 \, ^{+4.6} _{-15.1}$ & $ -26.7 \, ^{+17.3} _{-17.3}$	&  $-37.5 \, ^{+9.7} _{-17.0}$ 	&$-17.2 \, ^{+13.6} _{-15.1}$		&   $-32.2 \, ^{+10.1} _{-8.2}$  \\  
Fe\,{\sc xxvi} K$\alpha$ abs	& $E_{{\rm line}}$ (keV) 	&  $7.00 \, ^{+0.02}  _{-0.02}$	&   $7.00 \, ^{+0.03} _{-0.10}$ & $7.06 \, ^{+0.04}  _{-0.04}$	& $7.03 \, ^{+0.01}  _{-0.02}$	& $7.03 \, ^{+ 0.07}  _{-0.03}$		&   $7.05 \, ^{+0.02}  _{-0.02}$ \\  
line			&$\sigma$ (keV)           		& $<$0.05 			&   $<$0.04 			&   $<$0.05 			& $<$0.05 			&   $<$0.05 				&   $<$0.06  \\  
			&EW              (eV)    		& $-51.3 \, ^{+18.0}_{-18.5}$	&   $-36.1 \, ^{+2.5}_{-5.6}$ 	& $-57.0 \, ^{+20.9}_{-22.9}$	& $-56.5 \, ^{+6.5} _{-14.6}$ 	& $-39.6 \, ^{+19.5}_{-20.5}$		&   $-39.4 \, ^{+6.1}_{-7.6}$   \\  
Fe\,{\sc xxv} K$\beta$ abs	& $E_{{\rm line}}$ (keV) 	& $7.85$ fixed 			&   $ 7.90 \, ^{+0.01}_{-0.04}$ & $7.85$ fixed 			& $ 7.87 \, ^{+0.05} _{-0.03}$	& $ 7.84 \, ^{+0.04} _{-0.04}$		&   $ 7.85 \, ^{+0.05} _{-0.02}$ \\  
feature			&$\sigma$ (keV)           		& $<$0.05 			&   $<$0.05 			&  $<$0.05  			& $<$0.04 			&   $<$0.05 				&   $<$0.05  \\  
			&EW         (eV)    	 		& $-28.1 \, ^{+28.1} _{-30.8}$ 	& $ -21.4 \, ^{+5.1} _{-6.7}$	& $ -64.0 \, ^{+33.8} _{-35.6}$ & $ -29.4 ^{+8.1} _{-6.4}$	& $ -73.3 ^{+29.6} _{-30.7}$ 		&   $ -29.3 ^{+3.6} _{-8.6}$ \\  
Fe\,{\sc xxvi} K$\beta$ abs	& $E_{{\rm line}}$ (keV) 	& 8.26 fixed 			&   $8.20 \, ^{+0.07} _{-0.03}$ & 8.26 fixed 			& $8.25 \, ^{+0.05}_{- 0.03}$	& $8.11 \, ^{+0.06} _{-0.01}$		&   $8.23 \, ^{+0.07} _{- 0.03}$  \\  
line			&$\sigma$ (keV)           		& $<$0.05 			&   $<$0.05 			&   $<$0.05  			& $<$0.05 			&   $<$0.05 				&   $<$0.06  \\  
			&EW         (eV)    	 		& $>$$-41.1$ 			&   $ -25.0 \, ^{+6.0} _{-7.0}$ & $-15.7 \, ^{+15.7} _{-56.6}$ 	& $-27.3 \, ^{+8.4} _{-9.6}$  	& $-69.1 \, ^{+36.5} _{-37.9}$		&   $-19.3 \, ^{+6.9} _{-10.4}$ \\  
Fe\,{\sc xxv} edge	& $E_{{\rm line}}$ (keV) 		& 8.83 fixed 			&   8.83 fixed 			& 8.83 fixed 			& 8.83 fixed			& 8.83 fixed				&   8.83 fixed  \\  
			&$\tau$             			& $<$0.40 			&   $0.19 \, ^{+0.02} _{-0.02}$ &   $<$0.42  			& $0.14 \, ^{+0.05} _{-0.08}$	&   $<$0.32 				&   $0.28 \, ^{+0.04} _{-0.02}$  \\  
Fe\,{\sc xxvi} edge	& $E_{{\rm line}}$ (keV) 		& 9.28 fixed			&   9.28 fixed			& 9.28 fixed 			& 9.28 fixed			& 9.28 fixed				&   9.28 fixed	  \\  
			&$\tau$             			& $<$0.22 			&   $<$0.04 			&   $<$0.50  			& $<$0.03 			&   $<$0.20 				&   $<$0.01  \\  
Fe broad emission	& $E_{{\rm line}}$ (keV) 		& $6.24 \, ^{+0.27}_{-0.24}$ 	&    $6.80 \, ^{+0.13}_{-0.09}$	&  $6.80$   fixed		& $ 6.85 \, ^{+0.11} _{-0.05}$	&  $6.80$   fixed 			&   $ 6.77 \, ^{+0.02} _{-0.09}$\\  
line			&$\sigma$ (keV)           		& $0.8 \pm{0.2} $  		&  $0.37 \, ^{+0.29}_{-0.11}$	&  $0.3$   fixed 		& $ 0.35 ^{+0.14} _{-0.08}$ 	&  $0.3$   fixed			&   $ 0.26 ^{+0.13} _{-0.04}$  \\  
			&EW         (eV)    	 		& $260 \, ^{+270} _{-160}$ 	&    $91 \, ^{+9}_{-30}$	&  $<$130 			& $ 96 ^{+27} _{-18}$		& $ 170 ^{+90} _{-70}$			&   $ 141 ^{+30} _{-40}$ \\  
\noalign {\smallskip}                         
\hline  
\label{tab:lines}  
\end{tabular}  
\end{center}  
\end{table*}  
     
\section{Results}  
   
A total of 6 EPIC spectra of \src\ (3 MOS1 and 3 PN)   
from the on-axis observations reported  
in Table~1 were extracted.   
These were then   
rebinned to oversample the FWHM of the energy  
resolution by a factor 3 and to have additionally a minimum of 20  
counts per bin to allow use of the $\chi^2$ statistic.  
In order to account for systematic effects a 2\%  
uncertainty was added quadratically to each spectral bin.  
The photo-electric absorption cross sections of   
Morrison \& McCammon(\cite{m:83})   
are used throughout.   
All spectral uncertainties are given at 90\%  
confidence, unless indicated otherwise.  
  
The overall 2.0--10~keV continua were investigated using the same model 
as applied to the ASCA spectrum reported in Ueda et al.~(\cite{u:01}) 
consisting of absorbed blackbody and 
multicolor disk-blackbody (Mitsuda et al.~\cite{m:84}) components.  
Fits were made to all 6 spectra 
individually. Examination of the fit 
residuals shows that, especially in the 3~PN spectra, 
broad positive residuals are present near 
6.7~keV, requiring the addition of a broad Gaussian line in all 
3 fits with EWs of $\sim$100~eV and widths, $\sigma$, in the range 
$\sim$200--400~eV.   
The continuum parameters derived from all   
spectra are similar. For the multicolor disk-blackbody component 
$kT$ is in the range 0.5--0.7~keV while the $kT$ of the blackbody is 
1.1--1.2~keV. Both components are absorbed by an $N_{\rm H}$, of 
$3.1-3.8 \, 10^{22}$~atom~cm$^{-2}$.  Without the addition of the 
emission and absorption features discussed below the fit quality 
is rather poor with a \rchisq\ of 3.5--4.6.  
The addition of emission and absorption lines 
(modeled by Gaussians) successively decreases the \rchisq\ to 
2.0--2.4 (see below). 
The spectral parameters for the multicolor  
disk-blackbody 
and blackbody are consistent with
those obtained from the ASCA observation by Asai et al.\ (\cite{a:00}).

Two strong absorption features are observed in all 6 spectra at around  
6.7 and 7.0~keV (see Table~2).  
These features are consistent with being due to resonant  
K shell absorption from highly ionized iron Fe\,{\sc xxv}  
and Fe\,{\sc xxvi},   
respectively. The deeper 7.0~keV feature is almost certainly 
the absorption feature observed from \src\ by Ueda et al.~(\cite{u:01}) 
using ASCA. The addition of both features is highly significant. 
Including the 7.0~keV line reduces the \rchisq\ by 
0.6--0.8 while the addition of the  
6.7~keV feature reduces the \rchisq\ by 0.3--0.6. 
Two other absorption features at around 7.8 and 8.2~keV 
are also detected 
here for the first time in \src\ in all the PN spectra, while in the MOS1   
spectra they are   
visible only in one observation (Obs.~ID~1001), consistent with the  
lower effective  
area of this instrument, compared to the PN.  
The addition of these features reduces the \rchisq\ by 0.15--0.3 
and 0.04--0.07 for the 7.8 and 8.2~keV features, respectively. 
In both the MOS1 and PN spectra an absorption feature  
at $\sim$4.1~keV is also evident. This   
is consistent with being due to resonant K$\alpha$ 
absorption from H-like calcium (Ca\,{\sc xx}).  
All the detected absorption lines and their parameters are listed  
in Table~\ref{tab:lines}. We note that there is a strong interplay 
between the parameters of the broad emission line and the EWs of  
the two strong absorption features since these features cut strongly 
into the emission line.  
This is most clearly seen in  
Fig.~\ref{fig:spec} which shows the best-fit PN spectra together with 
expanded plots of the spectral regions that include the Fe features.

Line   energies measured using the PN Timing Mode spectra   
were expected to be slightly higher than their ``true'' values  
due to uncertainties in the calibration of Timing Mode.  
However, a comparison of the energies  
derived from the MOS1 and PN spectra for the two   
strongest absorption lines (Fe\,{\sc xxvi} and Fe\,{\sc xxv} K$\alpha$)  
shows that they  
are consistent at 90\% confidence. We therefore 
present in Table~\ref{tab:lines} the PN energies without 
any   adjustment.   
The identification of the fainter   
absorption features at 7.8 and 8.2~keV,   
is uncertain.  
They may be tentatively identified with Fe\,{\sc xxv} (7.88~keV)  
and Fe\,{\sc xxvi} K$\beta$ (8.26~keV),  
{\it or} with  Ni\,{\sc xxvii} (7.80~keV) and Ni\,{\sc xxviii}  
K$\alpha$ (8.10~keV), respectively     
(e.g., Kotani et al.~\cite{k:00}).   
 
Close inspection of the residual plots (Fig.\ \ref{fig:spec}) reveals 
the presence of an emission feature near 8.5~keV. A fit to this 
feature with a Gaussian model gives an energy of $8.51 \pm 0.05$~keV and an 
EW in the range 25--35 eV. Since this energy is close to those of known 
PN background features due to Zn K$\alpha$ and Cu 
K$\beta$ (an even stronger line is present in the background spectrum 
near 8.05~keV from Cu K$\alpha$) we investigated whether this feature 
could originate from the background.  
In the PN Timing Mode, data from the on-axis CCD are continually 
clocked across the target region and 
data are extracted 
with   only one dimension of spatial information.
We searched all the publically available XMM-Newton observations
but were unable to find any blank field
Timing Mode data from which to extract a background spectrum.
Since the internal background is spatially very   
structured, 
particularly in the K- fluorescent lines of metal components of the camera 
body, we 
constructed a background spectrum from the identical region covered by 
Timing Mode from Imaging Mode data, taking care to apply the appropriate 
exposure time correction. 
From this spectrum we 
find that the observed 8.5~keV feature is a factor 100 
more intense than the   background features.
We therefore conclude that the observed feature is not 
a background artifact.  
The energy of the emission 
feature does not correspond to 
any prominent emission line from an abundant element.  
However, it is just below that  
of the He-like Fe edge at 8.83~keV. 
We therefore attempted to model this feature as an absorption edge. 
The inclusion of an edge at 8.83~keV accounts for the excess around 8.5~keV 
and decreases the \rchisq\ to 1.6--1.7. 
The best-fit parameters for all the discrete features can 
be found in Table~\ref{tab:lines}, where the upper limits  
to the presence of a H-like Fe edge (at 9.28~keV) are also reported.  
A narrow emission feature near 6.4 keV was reported by Ueda et al.\ 
(\cite{u:01}). Addition of this narrow feature with an energy close to 
6.4 keV resulted in a decrease of the \chisq\  of 4--7, which   
is not significant.  
 
We note that the parameters derived for the broad emission 
feature from the PN and MOS1 are inconsistent. 
The energy derived from the 
MOS is lower, while the width of the line is significantly 
broader. This is not caused by the fitting method (e.g.\ a local 
minimum) since the addition of the line with the PN-derived  
parameters to the 
MOS spectral model results in a clearly worse \chisq\ 
($\Delta$\chisq$>$100), and the difference between the model and 
the MOS data is up to 
10--15\%, i.e.\ larger than the calibration uncertainty. 
Since a large fraction of the line emission is removed 
by the narrow absorption components it  is difficult to 
reliably constrain the emission component. We therefore await an 
improved EPIC calibration before the parameters of the broad emission 
line can be accurately determined.

\section{Discussion}  
\label{sect:discussion}  
  
There is a growing evidence that narrow absorption lines from highly ionized  
material are a common feature of LMXRB spectra.  
They were first detected from the superluminal  
jet sources GRO\,J1655$-$40 (Ueda et al.~\cite{u:98};   
Yamaoka et al.~\cite{y:01}) and GRS\,1915+105 (Kotani et al.~\cite{k:00};  
Lee et al.~\cite{l:01}) with ASCA.   
Absorption lines due to Fe\,{\sc xxv} and Fe\,{\sc xxvi}   
were observed from GRO\,J1655$-$40 and    
did not show any obvious dependence of their EWs on orbital phase.  
Both X--ray and optical observations indicate that 
GRO\,J1655$-$40 is most probably a binary system viewed at 
a high inclination angle of $\sim$$70\degmark$ (Kuulkers et al.~\cite{k:98};
van der Hooft et al. \cite{vdh:97}; Greene et al. \cite{gr:01}).
ASCA observations of GRS\,1915+105  
revealed, in addition, absorption features due to Ca\,{\sc xx},  
Ni\,{\sc xxvii} and Ni\,{\sc xxviii}.   
This source has been observed also with the  
{\it Chandra} HETGS which detected   
absorption edges of Fe, Si, Mg, and S  
as well as resonant absorption features from  
Fe\,{\sc xxv} and Fe\,{\sc xxvi} and possibly Ca\,{\sc xx}  
(Lee et al.~\cite{l:01}).  
The recent discovery of similar narrow iron absorption  features   
in X--ray binaries  
containing neutron stars (GX\thinspace13+1, Ueda et al.~\cite{u:01};   
MXB\thinspace1658$-$298, Sidoli et al.~\cite{si:01} and 
X\,1624-490, Parmar et al.~\cite{p:01}),  
rules out the possibility that such features are  
peculiar to superluminal jet sources and related in some way to the jet  
formation mechanism.  
{\it Chandra} HETGS observations have also convincingly detected 
narrow absorption features due to Ne\,{\sc ix} and Fe\,{\sc  xviii} 
from the transient
black hole candidate XTE\,J1650$-$500 (Miller et al. \cite{m:02}). 
These lines are may be due to absorption in an ionized accretion 
disk atmosphere or wind.
   
The XMM-Newton observations reported here reveal a much more 
complex range of emission and absorption components from \src\ than 
even the ASCA analysis of Ueda et al.~(\cite{u:01}) had found.  
In addition  
to the Fe~{\sc xxvi} K$\alpha$ absorption feature  
additional narrow absorption features which may be 
identified with Ca~{\sc xx} and Fe~{\sc xxv} K$\alpha$ and with 
Fe~{\sc xxv} and Fe~{\sc xxvi} K$\beta$ are evident in the XMM-Newton 
spectra.  
In addition a deep edge at 8.83~keV is clearly required.  
We 
note that the feature at $7.61 \pm 0.13$~keV which Ueda et al.~(\cite{u:01}) 
model as an edge is more clearly seen to be a line absorption feature 
in the XMM-Newton PN spectra (which we identify with Fe~{\sc xxv}  
K$\beta$).    
It is interesting to speculate as to why a deep ($\tau \sim 0.2$)  
edge from Fe~{\sc xxv} is evident whereas the PN upper-limits to the  
optical depth 
for an Fe~{\sc xxvi} edge at 9.28~keV of $\approxlt$0.04 exclude a  
similar 
feature. This is in contrast to the corresponding absorption lines, where 
the EW of the Fe~{\sc xxvi} transitions exceeds that of Fe~{\sc xxv} in all 
the spectra.  
  
\begin{figure}  
   
\includegraphics[width=8.5cm,angle=0]{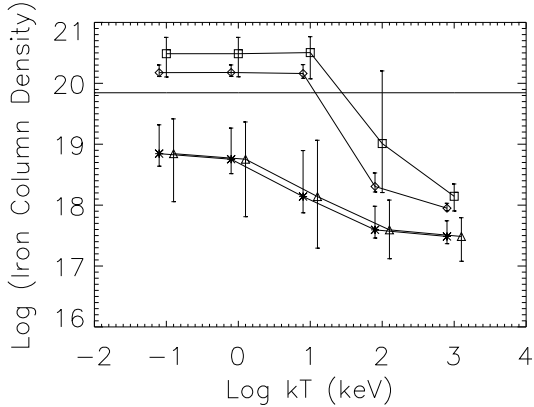}  
  \caption[]{Fe column densities for the two K$\alpha$ absorption lines   
  from the first (ID.~0101) MOS1 (squares are for Fe\,{\sc xxvi},  
  triangles for  Fe\,{\sc xxv}) and PN observations  
  (diamonds are for Fe\,{\sc xxvi},  
  stars for  Fe\,{\sc xxv}). For the other two observations  
  (ID.~0901 and ID.~1001) similar 
  results have been obtained.  
  The horizontal line marks the upper limit to the Fe column density  
  corresponding to a Thomson optical depth equal to unity.  
  A small horizontal shift has been applied for clarity. 
  Column densities are in units of cm$^{-2}$ 
  }  
 \label{fig:n_ironlines}   
\end{figure}

\begin{figure}  
  \includegraphics[width=8.5cm,angle=0]{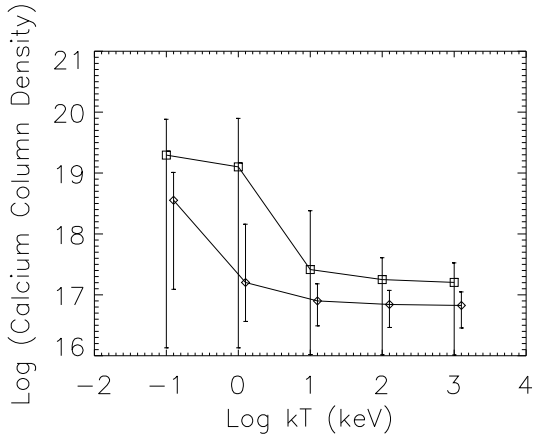} 
  \caption[]{Results of the curve of growth analysis for the  Ca\,{\sc xx}   
  line, using only results from the first observation (Obs.~ID~0101).  
  Similar values are obtained from the other two observations.  
  Squares corresponds to PN results, diamonds to MOS1.   
   A small horizontal shift has been applied for clarity.  
   Column densities are in units of cm$^{-2}$  
   }  
 \label{fig:n_caline}   
\end{figure}

In order to convert the observed EWs of the lines into ion  
column densities, we assumed the curve of growth analysis performed  
by Kotani et al.~(\cite{k:00}).  
In this way, the column densities of the lines can be estimated for a range  
of kinematic temperatures of the absorbing material, where  
this temperature includes contributions from thermal   
motions as well as any bulk motions or turbulence.  
An Fe abundance of $4.7 \, 10^{-5}$ was assumed  
(Anders \& Grevesse~\cite{a:89}).  
The  column densities for  Fe\,{\sc xxv} and Fe\,{\sc xxvi}, calculated  
for a range of assumed temperatures (0.1, 1, 10, 100 and 1000 keV),  
are shown in Fig.~\ref{fig:n_ironlines} for only one observation, for clarity.   
Indeed, similar EWs have been measured in all the three observations.  
A lower kinematic temperature requires a higher ion  
column density, and hence a higher \nh.   
A constraint on the absorbing Fe column density can be derived from the  
fact that the plasma should be optically thin to Thomson scattering   
(\nh\ $ < 1.5 \, 10^{24}$~atom~cm$^{-2}$),   
otherwise the absorption lines would   
be strongly diminished.   
Assuming a solar Fe abundance, implies an Fe column density  
$<$$7 \, 10^{19}$~cm$^{-2}$, which  
translates into a temperature $\approxgt$10~keV.  
The upper limits to the velocity widths of the two narrow lines  
of $<$1800~km~s$^{-1}$ (Fe\,{\sc xxv})   
and $<$1900~km~s$^{-1}$ (Fe\,{\sc xxvi}),  
provide not very constraining upper limits to the $kT$  
of 600--650~keV.  
The EWs of the highly ionized Fe K lines (and thus the ions column  
densities)   
are similar to those measured  
in the eclipsing, dipping LMXRB MXB\thinspace1658$-$298    
(Sidoli et al.~\cite{si:01}) with XMM-Newton, and a factor $\sim$3   
higher than that observed   
in the LMXRB X\,1624--490 (Parmar et al.~\cite{p:01}).  
The results of the curve of growth analysis for the Ca\,{\sc xx} line   
are shown  
in Fig.~\ref{fig:n_caline}. The ionized calcium column density  
ranges from 10$^{16}$~cm$^{-2}$ to 10$^{20}$~cm$^{-2}$,  
depending on the kinematic temperature.  
No useful constraints come from the upper limit to the   
velocity width of $<$3700~km~s$^{-1}$,  
that translates into a temperature of $\approxlt$1900~keV.

The line centroids of the Fe\,{\sc xxv} and Fe\,{\sc xxvi}  
K$\alpha$  do not show evidence for any velocity 
shifts, except in the second observation (Obs.~ID~0901) where blue-shifts 
of $-2700 \pm{2200}$~km~s$^{-1}$ (Fe\,{\sc xxv} K$\alpha$)  
and $-3900 \pm{1700}$~km~s$^{-1}$ (Fe\,{\sc xxvi} K$\alpha$) 
are evident.  
In this same observation, a blue-shift in the Ca\,{\sc xx} 
feature of $-5100 \pm {2200}$~km~s$^{-1}$ is also detected. 
For most observations, the MOS energy calibration is correct to within 
$\sim$5~eV, or a velocity shift of $\sim$220~km~s$^{-1}$ at 7~keV and
$\sim$400~km~s$^{-1}$ at 4~keV. 
Individual observations have been analysed where the Mn 
K$\alpha$ internal calibration line appears to be discrepant by 
$\sim$30~eV, but these are all associated with anomalous instrument 
temperatures resulting from spacecraft irregularities. Such 
conditions did not pertain for the observations reported here. 
The absorption lines at higher energy cannot be unambiguously identified. 
These features are only clearly detected in the MOS1 
in the third observation (Obs.~ID.~1001). 
The measured line energy of 7.84$\pm{0.04}$~keV can be associated 
with {\it both} the    
Fe\,{\sc xxv} K$\beta$ (at 7.88~keV) and the 
 Ni\,{\sc xxvii} K$\alpha$ (at 7.80~keV) feature. 
The absorption line at $\sim$8.1~keV can be similarly associated 
with either an Fe\,{\sc xxvi} K$\beta$ transition (at 8.212~keV) or 
with a Ni\,{\sc xxviii} K$\alpha$ feature at 8.034~keV. 
  
Since the orbital period of \src\ has not been confirmed, 
we cannot reliably estimate a geometry and size for the   
absorbing plasma around the central  
X--ray source.  
If the 25~day period is confirmed to  
be orbital, we note that the \xmm\  
observations reported here only cover a small fraction ($\sim$10\%) 
of an orbital cycle.  
All we can say is that all the narrow absorption lines are consistent   
with being  
present with the same EW in all analyzed spectra.   
This implies that the absorbing plasma is stable in the line of sight   
over timescales of few days suggesting a stable geometry, as in the case of   
MXB\thinspace1658$-$298 and X\,1624--490, and excluding the possibility of  
absorbing clouds temporarily passing  
in front of the central source producing the absorption.  
Additionally, the fact that the EWs of the Fe\,{\sc xxvi} line measured   
here are consistent  
with that observed with ASCA (Ueda et al.~\cite{u:01}) a few years ago,  
again  
supports this conclusion.  
  
\begin{acknowledgements}  
Based on observations obtained with XMM-Newton, an ESA science mission  
with instruments and contributions directly funded by ESA member states  
and the USA (NASA).    
We thank  
T.~Kotani for making  
his curve of growth software available.  
  
\end{acknowledgements}

\end{document}